\documentclass[prl,aps,twocolumn,superscriptaddress,floatfix,showpacs,10pt]{revtex4-1}
\usepackage{graphicx,amsfonts,amssymb,amsmath,hyperref,dsfont}

\newif\ifhyper
\hypertrue
\ifhyper
\hypersetup{
   citecolor = {green},
   colorlinks = {true}, 
   urlcolor = {blue} 
}
\fi
 \newcommand{\aoo}[1]{\alpha_{\bf{#1}}^{\phantom{\dagger }}}
 \newcommand{\ad}[1]{\alpha_{\bf{#1}}^{\dagger} } 

\newcommand{\beq}{\begin{equation}}
\newcommand{\eeq}{\end{equation}}
\newcommand{\beqa}{\begin{eqnarray}}
\newcommand{\eeqa}{\end{eqnarray}}

\newcommand{\bo}[1]{\beta_{\bf{#1}}^{\phantom{\dagger}} } 
\newcommand{\bd}[1]{\beta_{\bf{#1}}^{\dagger} } 
\newcommand{\dG}[1]{\delta_G \left({\bf{#1}}\right) } 
\newcommand{\w}[1]{\omega\left({\bf #1}\right)} 
\newcommand{\coeff}[2]{C_{#1}\left( \bf {#2}  \right)} 
\newcommand{\coeffgamma}[1]{\Gamma\left( \bf {#1}  \right)} 
\newcommand{\uminus}{\scalebox{0.5}[1.0]{\( - \)}}

\def\Longarrow{\protect\@lra}
\def\@lra{\relbar\joinrel\relbar\joinrel\relbar\joinrel%
          \relbar\joinrel\rightarrow}

\def\be{\begin{equation}}       \def\ee{\end{equation}}
\def\bea{\begin{eqnarray}}      \def\eea{\end{eqnarray}}
\def\bes{\begin{subequations}}  \def\ees{\end{subequations}}

\def\dag{\dagger}
\def\non{\nonumber}

\def\k{{\bf k}}

\begin{document}

\title{Roton minimum as fingerprint of magnon-Higgs scattering in ordered quantum antiferromagnets}

\author{M.~Powalski}
\email{michael.powalski@tu-dortmund.de}
\affiliation{Lehrstuhl f\"ur Theoretische Physik 1, TU Dortmund, Germany}
\author{G.~S. Uhrig}
\email{goetz.uhrig@tu-dortmund.de}
\affiliation{Lehrstuhl f\"ur Theoretische Physik 1, TU Dortmund, Germany}
\author{K.~P.~Schmidt}
\email{kai.schmidt@tu-dortmund.de}
\affiliation{Lehrstuhl f\"ur Theoretische Physik 1, TU Dortmund, Germany}

\date{\rm\today}

\begin{abstract}
 A quantitative description of magnons in long-range ordered quantum antiferromagnets  is presented which is consistent from low to high energies. It is illustrated for the generic $S=1/2$ Heisenberg model on the square lattice. The approach is based
on a continuous similarity transformation in momentum space using the scaling dimension
as truncation criterion. Evidence is found for significant magnon-magnon
attraction inducing a Higgs resonance. The high-energy roton 
minimum in the magnon dispersion appears to be induced by strong magnon-Higgs scattering.  
\end{abstract}



\maketitle

%
%
\emph{Introduction ---}
%
%
The characterization of elementary excitations is of fundamental 
importance in condensed matter physics. Especially
 strongly correlated systems are known to host a large range of 
fascinating collective behavior leading to a zoo of exotic elementary 
 quasi-particles with integer or non-integer, i.e, fractional, quantum numbers.
The decay of integer excitations into fractional ones is called fractionalization.
Prominent examples are the fractional quantum Hall effect 
with fractional charge excitations \cite{laugh83} or the physics 
of magnetic monopoles in spin ice where spin north and south poles move
independently \cite{caste08}. 
Another arena of major importance is high-temperature superconductivity,
where the nature of magnetic excitations has been intensively debated. 
The key point is whether fractional spin excitations, 
spinons with $S=1/2$, are present and essential \cite{ander87}, or 
whether a description in terms of magnons or triplons with integer spin $S=1$ is 
appropriate \cite{knett01b}.

Theoretically, the physics of undoped cuprates is closely
 related to the antiferromagnetic $S=1/2$ Heisenberg model 
on the two-dimensional (2D) square lattice \cite{manou91}. 
This paradigmatic unfrustrated model 
is known to display long-range N\'eel order with finite sublattice 
magnetization at zero temperature \cite{reger88,chakr89}. The well-understood
low-energy excitations 
are gapless magnons as enforced by Goldstone's theorem \cite{auerb94}.

Remarkably, the magnetic excitations at high energies
are  understood less well. But they have come into focus recently due
to remarkable experimental progress in resolution \cite{chris07a,dalla15}
and in the direct excitation at high energies \cite{letac11}.
Conventional spin-wave (magnon) theory in powers of $1/S$ 
\cite{hamer92,weiho93,igara05,syrom10,uhrig13} fails 
to agree quantitatively with unbiased  quantum Monte Carlo (QMC) simulations 
\cite{sandv01} and high-order series expansions about the Ising limit 
\cite{sandv01,zheng05}. Even in third order in $1/S$, the approach based on magnons
yields only a weak dip of the dispersion of about $3\%$ at $(\pi,0)$ compared
to its value $(\pi/2,\pi/2)$  while series expansion and QMC yield about $10\%$.
Inelastic neutron scattering (INS) finds $7(1)\%$ in deuterated copper formate
\cite{chris07a,dalla15}.
In analogy to the similar feature in the dispersion of excitations in 
$^4$He \cite{vollh90b} we call this dip `roton minimum'.

The agreement of the INS data with the QMC and series results
for the roton minimum on the one hand and the strong deviation of
spin wave theory  on the other hand 
have revived the idea, see, e.g., Ref.\ \onlinecite{ho01}, 
that at high energies magnons are not the
appropriate elementary excitations and  that spinons 
provide a better description \cite{dalla15}. The magnons
at lower energies would result as confined states of spinon pairs
\cite{tang13} while the spinons are deconfined at least at certain
points in the Brillouin zone (BZ) at higher energies.

The ambiguity between excitations having fractional or integer spin 
arises in gapless systems because without gap it is
an ill-posed problem to determine the content of quasi-particles.
This was for instance seen in one-dimensional (1D) spin chains  \cite{schmi03c}. But in contrast to the 2D case, where magnons are the established
Goldstone modes above the long-range ordered ground state, 
the 1D Heisenberg chain is known to display fractional massless spinons  \cite{fadde81}. 
Consequently, a consistent determination of the full magnetic excitation spectrum for the
2D Heisenberg model from low to high energies
 is still missing, but highly desirable. 

This is the central goal of the present Letter. 
We provide a quantitative description of the low- and the high-energy part of the 
dispersion in terms of magnons. Technically, this is achieved by extending 
continuous unitary transformations to continuous similarity transformations (CST)
which are carried out in momentum space using the scaling dimensions of
operators as appropriate truncation criterion.  

Our results indicate a strong attractive magnon-magnon interaction.
This interaction induces a  Higgs resonance built from two-magnon states,
also addressed as longitudinal magnon \cite{lake00,ruegg08,varma15}. 
Thereby, significant weight in the three-magnon continuum 
is shifted towards lower energies and it can also be seen as magnon-Higgs continuum. 
These scattering states depress the single magnon states and
reduce their energy yielding the roton minimum. 
In this way, this dip in the dispersion is a fingerprint of strong scattering 
between magnons and the damped Higgs mode.

%
%
\emph{2D Heisenberg model ---}
%
%
%
We consider the Heisenberg quantum antiferromagnet on the square lattice
\begin{equation}
 \mathcal{H} = J \sum_{\langle i,j\rangle} \vec{S}_{i}\cdot \vec{S}_{j}\quad,
 \label{eq:ham}
\end{equation}
%
where $\vec{S}_{i}$ denotes the vector operator of a spin $S=1/2$ at site $i$ 
and the sum runs over pairs of nearest neighbors. 
The ground state displays long-range N\'eel order. Consequently, the classical N\'eel state with spin up (spin down) on sublattice $A$ ($B$) serves as the appropriate reference state with respect to which bosonic operators
create deviations. We use the non-hermitian Dyson-Maleev representation 
\cite{dyson56a,malee57} as is standard in analytic higher order calculations
\cite{hamer92,weiho93,igara05,syrom10,uhrig13}. Its asset is that the Hamiltonian only comprises terms that are at most quartic in the bosonic operators. 

In this way, we obtain an initial Hamiltonian first in real space and
by Fourier transformation in momentum space.  A Bogoliubov transformation introducing 
$\alpha_{\bf k}^{(\dag)}$ and $\beta_{\bf k}^{(\dag)}$ as new bosonic operators
yields \cite{uhrig13,suppl15}
 \begin{equation}
   \mathcal{H} = E_0 + \sum_{\k}\omega_{\k} 
	( \alpha^\dagger_{\k}\alpha^{\phantom{\dagger}}_{\k}+ 
	\beta^\dagger_{\k}\beta^{\phantom{\dagger}}_{\k})+\Gamma+\mathcal{V}.
	\label{eq:ham_initial}
\end{equation}
The momenta $\k$ are taken from the magnetic BZ containing $N$
points given by the diamond with the four edges $(\pm\pi,0)$ and $(0,\pm\pi)$.
 The coefficients \mbox{$E_0/N=-2J (S^2+AS +A^2/4)$} and 
\mbox{$\omega_{\k}=2J(2S+A)\sqrt{1-\gamma_{\k}^2}$} with 
\mbox{$\gamma_{\k}=(\cos k_x+\cos k_y)/2$} and 
$A:=\frac{2}{N}\sum_\k(1-\sqrt{1-\gamma_{\k}^2})\approx 0.158$ correspond to the ground-state energy per site and the one-magnon dispersion in next-to-leading order spin-wave theory 
\cite{hamer92,uhrig13}. The Bogoliuov part $\Gamma:=\Gamma_0^2+\Gamma_2^0$ is bilinear; 
$\Gamma_n^m$ consists of $m$ creation and $n$ annihilation operators.
 The quartic terms are split according to \mbox{$\mathcal{V}:=
\mathcal{V}_2^2+\mathcal{V}_1^3+\mathcal{V}_3^1+\mathcal{V}_0^4+\mathcal{V}_4^0$}
with the same pattern of bosonic operators.
By construction, $\Gamma=0$ in \eqref{eq:ham_initial} once the Bogoliubov transformation
is carried out self-consistently. We define the part with
more (less) creation than annihilation operators by 
$\mathcal{H}^+:=\Gamma_0^2+\mathcal{V}_1^3+ \mathcal{V}_0^4$ 
($\mathcal{H}^-:=\Gamma_2^0+\mathcal{V}_3^1+ \mathcal{V}_4^0$).

Our goal is to derive an effective Hamiltonian $\mathcal{H}_{\rm eff}$ which conserves the number of magnons in order to simplify the subsequent analysis. We achieve this by extending the approach of CUTs \cite{wegne94} to similarity transformations using the flow equation 
\mbox{$\partial_{\ell} \mathcal{H}(\ell) = \left[\eta(\ell),\mathcal{H}(\ell) \right]$}, where $\ell$ is a continuous auxiliary variable which labels the stage of the
CUT: $\mathcal{H}(\ell)$ has the same structure as in \eqref{eq:ham_initial},
but all coefficients acquire a dependence on $\ell$. Initially, one has 
\mbox{$\mathcal{H}(\ell=0)\ := \mathcal{H}$} and finally
\mbox{$\mathcal{H}_{\rm eff}:=\mathcal{H}(\ell=\infty)$}. 

We use the generator 
\mbox{$\eta(\ell):=\mathcal{H}^+(\ell)-\mathcal{H}^-(\ell) $} 
\cite{knett00a,fisch10a} leading to an effective 
Hamiltonian $\mathcal{H}_{\rm eff}$ that conserves the number of magnons. 
This is well-established for the converging flow of hermitian Hamiltonians and
carries over to $\mathcal{H}$ in \eqref{eq:ham_initial} because the 
Dyson-Maleev transformation \emph{does not mix} states with different magnon
number, but only attributes prefactors to them which spoil its unitarity.
Summarizing, one maps the complicated many-body problem to an effective few-body Hamiltonian being block-diagonal in the number of magnons. The one-magnon block in particular is diagonal in momentum space after the CST so that the dispersion 
$\tilde{\omega}_{\k}$ can be read off.

Computing $\left[\eta(\ell),\mathcal{H}(\ell) \right]$ on the right hand side
of the flow equation leads to a proliferating number of terms. One needs a
systematically controlled truncation scheme to close the flow equation.
So far, perturbative truncations have been used mostly 
\cite{knett00a,knett03a,kehre06,krull12}.
In real space approaches the spatial range of processes is a related truncation
criterion appropriate for gapped systems with finite
correlation length \cite{fisch10a,yang11a,krull12}. In 1D, the scaling
dimension was used in the operator product expansion of vertex operators 
\cite{kehre99,kehre01}. The criterion of the scaling dimension is particularly
suitable for gapless systems with infinite correlation length.
Hence, we apply this criterion for the 2D massless magnons.

We consider a generic term $\mathcal{T}$ 
\begin{equation}
\label{eq:scaling}
 \mathcal{T} \!\!= \!\!\int C_{\k_1\ldots\k_{n}}
\mathcal{O}^n_{\k_1\ldots\k_{n}} \delta(\k_1+\ldots+\k_n)
d^2\k_1\ldots d^2\k_n ,
\end{equation}
where $\mathcal{O}^n_{\k_1,\ldots\k_{n}}$ is a monomial of $n$ bosonic
operators of creation or annihilation type
\footnote{For the scaling one must 
consider the thermodynamic limit with 
continuous momenta as in \eqref{eq:scaling}. The actual 
calculations are done with discrete momenta.}.
If we rescale all momenta
$\k_i\rightarrow\lambda\k_i$ with $\lambda<1$ and assume that 
the coefficient is a homogeneous function 
$C_{\lambda\k_1\ldots\lambda\k_{n}} = \lambda^c C_{\k_1\ldots\k_{n}}$
the term $\mathcal{T}$ acquires a prefactor $\lambda^{n-2+c}$.
The scaling dimension is the exponent $d:=n-2+c$. It quantifies
the relevance of the term if we zoom to small energies
in the sense of renormalization. Smaller scaling dimensions stand for
more important terms.
If the coefficient $C$ is not a homogeneous function we determine its
dimensional contribution $c$ from the leading term in a series expansion in $\lambda$.

%
\begin{figure}[ht]
	\centering
		\includegraphics[width=\columnwidth]{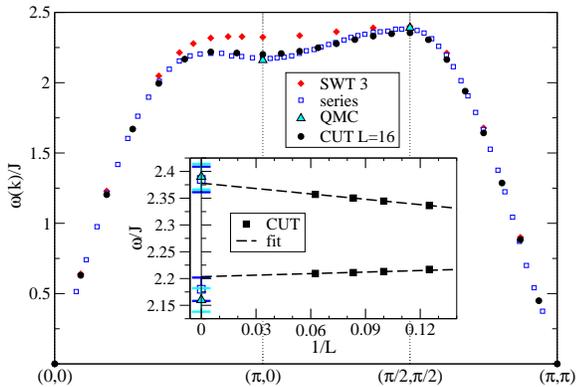}
	\caption{One-magnon dispersion $\omega({\bf k})$ in the magnetic BZ. 
	Circles depict CST data for $L=16$; diamonds,
	squares, and triangles show data from spin-wave theory \cite{syrom10}, 
	series expansion and QMC extrapolated to the thermodynamic limit 
	\cite{sandv01}. 
	{\it Inset}: Extrapolation of the dispersion at momenta 
	$(\pi,0)$ and $(\pi/2,\pi/2)$ (circles). The squares depict the series expansion and 
	the triangles up the QMC data \cite{sandv01}.}
	\label{Fig:dispersion}
\end{figure}

The bilinear kinetic energy 
$\omega_\k( \alpha^\dagger_{\k}\alpha^{\phantom{\dagger}}_{\k}+ 
	\beta^\dagger_{\k}\beta^{\phantom{\dagger}}_{\k})$ has scaling dimension
	$d=2-2+1=1$ because the dispersion is linear for small momenta. 
	Thus, this scaling dimension is the one to compare to; conventionally 
	it is called marginal. Terms with smaller dimension would be relevant,
	terms with larger dimension irrelevant. The Bogoliubov terms 
	$\Gamma=\sum_{\k}\Gamma_{\k} (\ell) 
	( \alpha^\dagger_{\k}\beta^{\dagger}_{-\k}+ {\rm h.c.})$
	are marginal because we find
	$\Gamma_{\k} (\ell) \propto |\k|$ as for the dispersion.
	The quartic terms in $\mathcal{V}$, however, have scaling dimension $2$
	because their prefactor is bounded for small momenta, i.e., $c=0$. 
	Most importantly, the hexatic three-particle terms with $n=6$ have
	scaling dimension $4$.
	
	Thus, the dispersion computed in scaling dimension $1$
	is the self-consistent mean-field result given below 
	\mbox{Eq.\ \eqref{eq:ham_initial}}. Here we improve this result by
	performing a fully self-similar CST for scaling dimension $2$. We make
	all coefficients of the Hamiltonian \eqref{eq:ham_initial} $\ell$-dependent, compute
	the commutators, normal-order the result, and discard the higher hexatic terms
	to deduce the differential flow equations for the coefficients. The flowing
	Hamiltonian, the generator and the differential equations are 
	given in the Supplement \cite{suppl15}.
  The differential equations are solved numerically by an adaptive Runge-Kutta 
	algorithm on finite systems with 
	$N=L^2$ points in the magnetic BZ up to $L=16$. 
We stop the flow at values of $\ell$ where the residual off-diagnality, i.e.,
a measure of the norm of the generator $\eta$ \cite{fisch10a}, has dropped below
$10^{-6}J$. The error due to the finite Runge-Kutta step size is approximately also $10^{-6}J$.
Furthermore, we checked that beyond this $\ell$ the ground-state energy and 
the dispersion do not change more than $10^{-9}J$
so that we consider them to be converged within good accuracy.

%
%
\emph{Low-energy properties ---}
%
%
At low energies, the magnons are the natural excitations 
so that we expect that our calculation 
agrees quantitatively with other results, e.g., QMC and series expansion
 \cite{sandv01}, which is indeed confirmed.  We find a ground-state energy per site 
\mbox{$\epsilon_0\equiv\tilde{E}_0/(2N)=-0.66939(3)$}
agreeing within $5\cdot 10^{-5}$ with $\epsilon_0^{\rm QMC}=-0.669437(5)$  from QMC \cite{sandv97b}. Recall that the variational Gutzwiller approach yields $\epsilon_0^{\rm Gutz}=-0.664$ corresponding to a gapped state with finite correlation length \cite{dalla15}.

The data for the spin-wave velocity $v$ agrees as well \cite{suppl15}.
It has been determined along two different directions in the BZ
yielding $v_1$ (along $(k,k)$) and $v_2$ (along $(k,0)$).
Their linearly extrapolated thermodynamic values coincide within the error bar yielding  
$v/J=1.675\pm0.025$. This value is in accord with $v^{\rm QMC}/J=1.673\pm0.007$ from QMC
 \cite{sandv97b}. Summarizing, the low-energy properties from the CST is quantitatively
consistent with the known properties of the Heisenberg model on a square lattice.

%
%
\emph{High-energy properties ---}
%
%
%
The complete one-magnon dispersion for $L=16$ is shown in Fig.~\ref{Fig:dispersion}
and compared to results from spin-wave theory in order $1/S^3$  
\cite{syrom10}, high-order series expansion, and QMC \cite{sandv01}. 
As discussed above, all approaches agree at low energies.

For the roton minimum at higher energies, this does not hold, see Fig.\ 
\ref{Fig:dispersion}. This is seen best in the dispersion difference between
momenta $(\pi,0)$ and $(\pi/2,\pi/2)$. The corresponding energies
are  extrapolated in $1/L$ in the inset of Fig.~\ref{Fig:dispersion}. The 
roton minimum represents a relative dip of (10.5$\pm$2.5)\% in QMC \cite{sandv01} and 
(9.5$\pm$2)\% for series expansions \cite{sandv01,zheng05} 
while spin-wave theory even in order 
$1/S^3$  \cite{syrom10} only yields \mbox{3.2\%}. Remarkably, the CST data at 
$L=16$ yields a sizable roton minimum of 7\%, and the extrapolation to $L=\infty$
yields 8\%.
In view of the uncertainties of all approaches, also the high-energy part of the dispersion including the roton minimum is quantitatively reproduced by the CST
in terms of magnons.

%
%
\emph{Discussion ---}
%
%
As long as there are processes linking 
states with a single elementary excitation, i.e., a quasi-particle, to 
states with two or three quasi-particles, the quasi-particles decay into
continua of two or three quasi-particles \cite{zhito13}. 
For collinear N\'eel order, the Hamiltonian allows only for decay into
three quasi-particles.
Generally, this hybridization lowers the energy of single quasi-particle states.
A well-studied example is the spin ladder with reflection
symmetry where the number of triplons (the elementary excitations) 
can only change by an even number \cite{schmi05b}. The triplon dispersion is pressed down by
the three-triplon continuum. In asymmetric spin ladders without 
reflection symmetry the single triplon states hybridize already 
with two-triplon continua. Due to the larger phase space of two-triplon
states this decay channel has a larger impact \cite{fisch10a,fisch11a,fisch11bb}
which is a general feature \cite{zhito13}. We conclude 
that one must understand the multi-magnon continuum
above the single magnon state in order to quantitatively assess the 
magnon dispersion in general and the roton minimum in particular.

In gapless systems such as the 2D Heisenberg model \eqref{eq:ham} the continua
start just above the single-magnon energies because multi-magnon
scattering states can be built from the single magnon and one or two magnons
arbitrarily close to  $\omega_{\bf 0}=0$. Thus, the crucial impact of the continua nearby in energy is
mainly driven by the low energy physics defined by terms of
low scaling dimension. 
Therefore, the scaling dimension is an appropriate criterion 
even for a calculation concerned with the high-energy roton minimum.

The energy lowering due to the hybridization with three-magnon states
is less effective as argued in the analogous system of
spin ladders. But an attractive interaction among the 
magnons shifts spectral weight to lower energies and enhances the impact
of the magnon decay. The marked roton minimum indicates that this
is a very important aspect. Indeed, the term $\mathcal{V}_2^2$ contains 
the attractive interaction $V^{(4)}$
 between the $\alpha$ and $\beta$ magnons
living on the $A$ ($S^z_\text{tot}=-1$) and the $B$ sublattice 
($S^z_\text{tot}=+1$), respectively \cite{suppl15}.

%
\begin{figure}[ht]
	\centering
		\includegraphics[width=0.99\columnwidth]{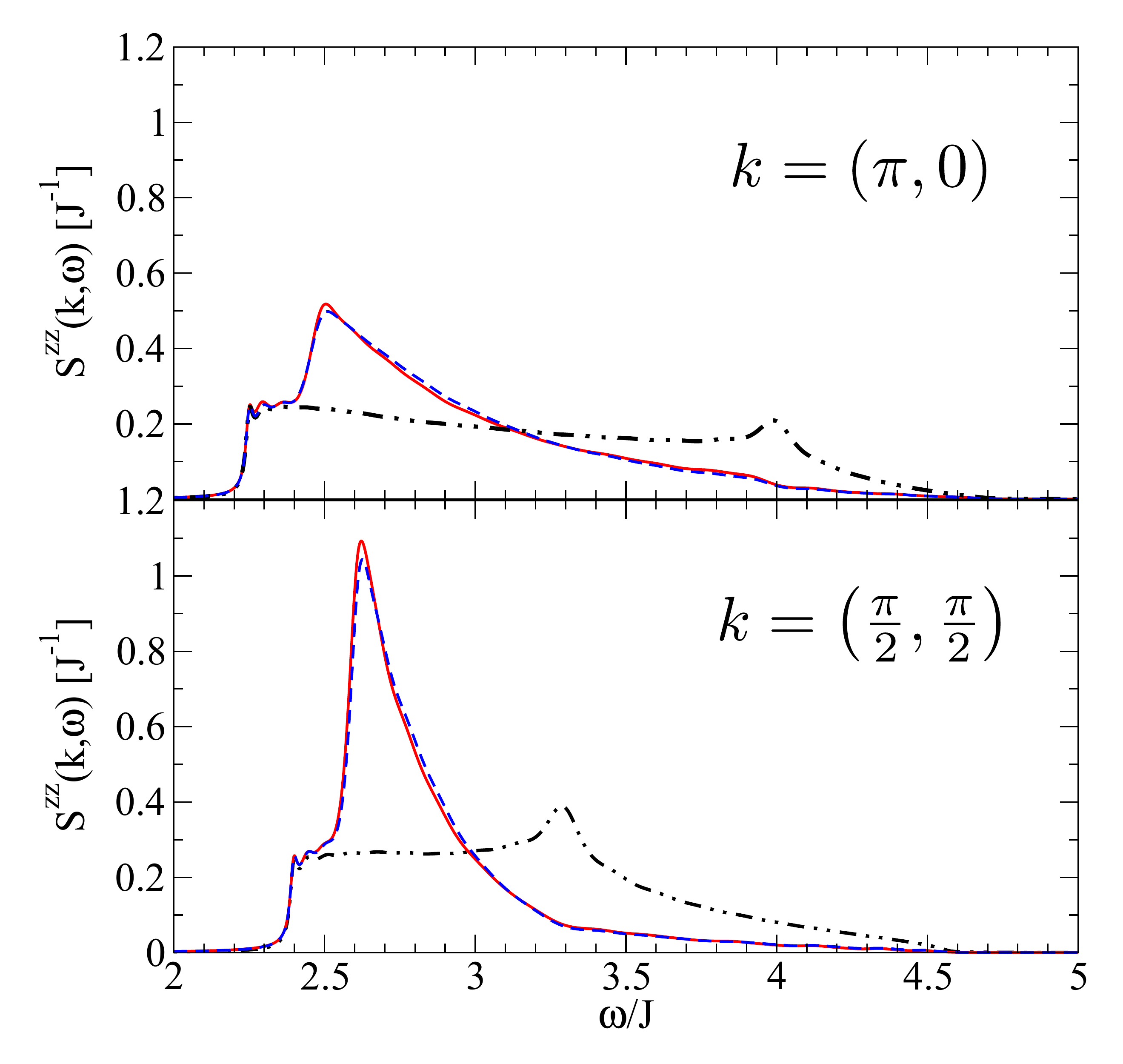}
	\caption{Two-magnon part of the 
	longitudinal dynamic structure factor $S^{zz}({\bf k},\omega)$. Dotted black line: without interaction, dashed blue line:  bare interaction, solid red line: renormalized interaction. 
	The curves are computed for $L=192$ using 
	the matrix elements of the $L=16$ system
	and broadened by $0.013J$. The dispersion
	is interpolated smoothly, the interaction in a piecewise 
	constant way.}
	\label{Fig:higgs}
\end{figure}

The crucial effects of the magnon-magnon attraction are
depicted in Fig.\ \ref{Fig:higgs} where we computed the
longitudinal dynamic structure factor $S^{zz}({\bf k},\omega)$
in the two-magnon $S^z_\text{tot}=0$
channel without and with (renormalized) interaction. 
The CST-renormalization of the observable is not included to
focus on the pronounced interaction effects. The attraction 
leads to a shift of spectral weight
to lower energies and, strikingly, to the formation of 
a resonance which we identify as the Higgs resonance.
The position of the resonance at $(\pi,0)$ is found to
be lower than the one at $(\pi/2,\pi/2)$. This is in line
with the argument that the single-magnon dispersion is pushed
to lower energies more strongly at $(\pi,0)$ than at $(\pi/2,\pi/2)$
leading to the roton minimum because due to the Higgs resonance
the decay of a single magnon not only comprises the decay into
three rather independent magnons, but also the
decay into two constituents:  magnon plus Higgs resonance.
In accord with previous studies 
\cite{schmi05b,fisch10a,fisch11a,fisch11bb,zhito13}
this is an efficient mechanism to lower the magnon dispersion and
to induce the roton minimum.

The renormalization of the interaction in the CST flow
enhances the effect of the attraction further. Though
this effect is small in absolute terms it makes itself
felt in the size of the roton dip. 
If we switch off the flow of the particle conserving
interactions, i.e., without renormalization of the magnon-magnon interaction,
the dip of the roton minimum is reduced to $5\%$.
This also explains why the third order calculation in $1/S$ 
yielded only a small dip although some interaction effect
was included (diagram (k) in Ref.\ \cite{syrom10}).
Still, the roton minimum is insufficiently captured.
Note that the renormalization of the interaction can
be as large as $50\%$ for certain momenta \cite{suppl15}.

We emphasize that our line shapes in the longitudinal channel
agree remarkably well with the experimental data,
see Figs.\ 2c and 2g in Ref.\ \cite{dalla15}. This holds for 
the position of the Higgs resonances and for the relative heights.
Investigating the
experimental transverse dynamic structure factor (Figs.\ 2b and 2f in
Ref.\ \cite{dalla15}) one finds pronounced peaks at $(\pi,0)$ and at
$(\pi/2,\pi/2)$. We attribute both of them to magnons and the smaller 
weight at $(\pi,0)$, combined with a continuum tail, to strong magnon-Higgs 
scattering. This also shifts the dispersion to
lower energies at $(\pi,0)$. Hence, the roton minimum in the 
magnon dispersion is a fingerprint of the Higgs mode and
of strong magnon-Higgs scattering.

%
%
\emph{Summary ---}
For the $S=1/2$ Heisenberg antiferromagnet on the
square lattice, this letter provides a fully
consistent and quantitative picture in terms of magnons
which describes the low- and the high-energy physics 
on equal footing. It is achieved by extending the
approach of continous unitary transformations to continuous
similarity transformations. In momentum space, 
the flow of the renormalized couplings is closed by the 
truncation at the level of terms with scaling dimension 2.

The striking agreement of our findings
with the experimental results \cite{dalla15}
underlines the validity of the magnon quasiparticle picture
for long-range ordered quantum magnets at all energies.
In this way, the importance of a significant attractive interaction
between magnons of different $S^z_\text{tot}$ is established.
This interaction induces a resonance corresponding to the
longitudinal magnon or Higgs resonance of the symmetry broken phase.
The decay of a  magnon into three magnons is strongly enhanced
because it is effectively a decay into a magnon and a Higgs resonance.
The roton minimum is the fingerprint of this mechanism.

Further studies should address other response functions 
providing predictions to spectroscopic experiments.
We highlight that the developed approach is applicable to
a humongous variety of long-range ordered phases. It will help to identify
their dynamical properties and eventually their instabilities and ensuing
transitions towards other exotic quantum phases.

\acknowledgments
This work was supported by the Helmholtz Virtual Institute 
``New states of matter and their excitations'' and by the Cusanuswerk (MP). 
We thank N.\ Christensen, H.\ R\o{}nnow, A.\ Sandvik, R.\ Singh, and A.\
 Syromyatnikov for fruitful discussions and exchange of data.

\bibliographystyle{apsrev4-1}

\onecolumngrid

\section*{Supplementary material}    

This supplementary materials gives all specific informations concerning 
the extrapolation of the spin wave velocity and the renormalization
of the interaction. In addition, details of the  CST performed in the main body of the manuscript are given.

\subsection{Spin wave velocity and magnon-magnon interaction}

%
\begin{figure}[ht]
	\centering
		\includegraphics[width=0.6\columnwidth]{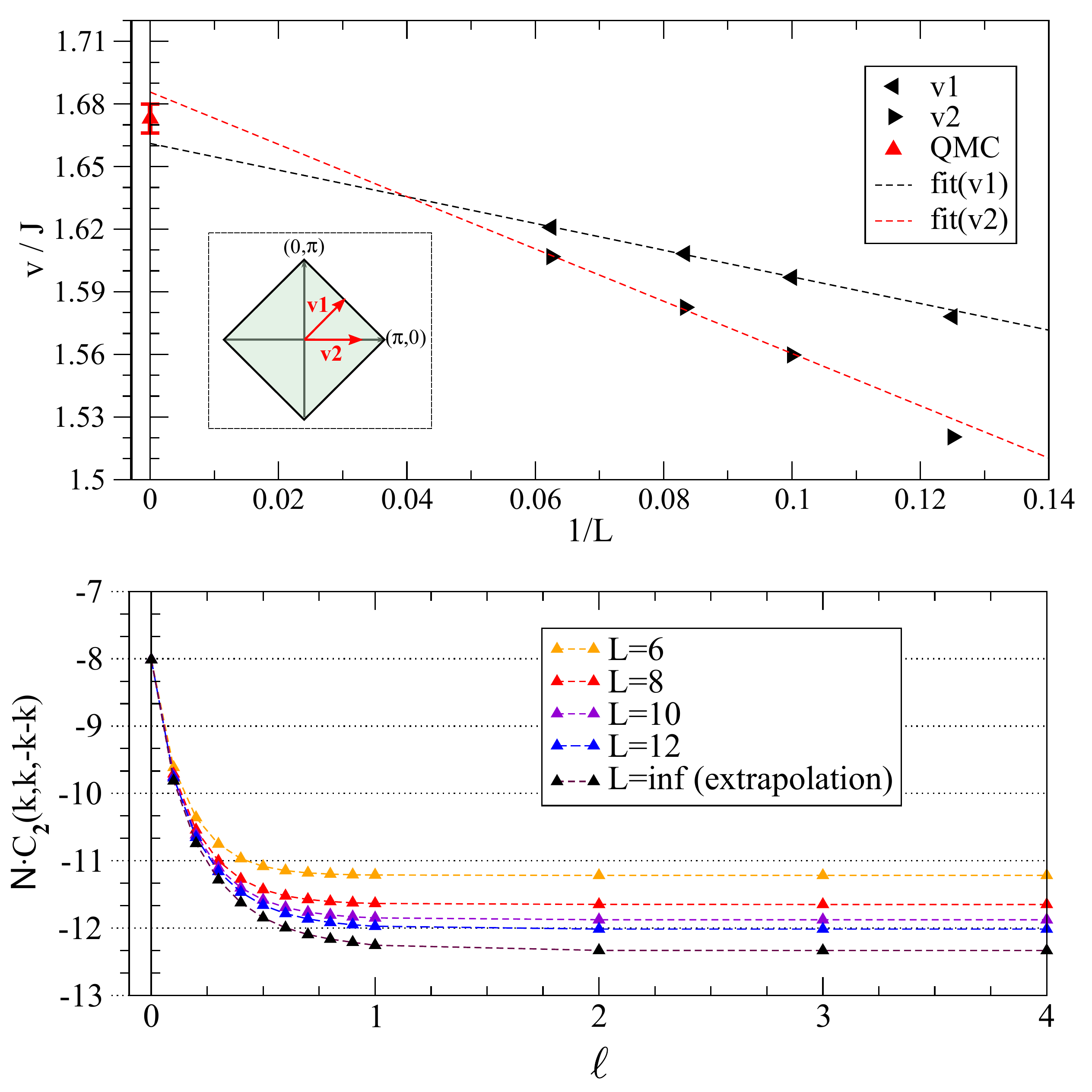}
	\caption{Extrapolation of the results in $1/L$ and comparison to other findings. All lines are linear fits. {\it Upper panel}: Spin-wave velocity $v$ deduced from two different directions $v_1$ (triangles left) and $v_2$ (triangles right) shown in the magnetic 
	Brillioun zone (inset). {\it Lower panel}: Renormalization of the attractive interaction of two magnons for smallest, non-zero $\k$ values along direction $v_1$ (see inset upper panel)
		vs.\ the flow parameter $\ell$; results for direction $v_2$ are almost the same.}
	\label{Fig:dip_velocity}
\end{figure}

Our data for the spin-wave velocity $v$ agrees well
with previous results as shown in the upper panel of Fig.~\ref{Fig:dip_velocity}.
It has been determined along two different directions in the BZ
yielding $v_1$ and $v_2$, see upper panel of Fig.~\ref{Fig:dip_velocity}.
Their linearly extrapolated thermodynamic values coincide within the error bar yielding $v/J=1.675\pm0.025$. This value is in accord with 
$v^{\rm QMC}/J=1.673\pm0.007$ from QMC  \cite{sandv97b}. 

In the lower panel, we illustrate that the CST procedure
renormalizes the interaction sizably. For the smallest non-vanishing
momenta the interaction is attractive and renormalized by up to
$50\%$. Averaged over all interaction coefficients the renormalization
is of the order of $10\%$. Note in addition that not all terms
of the interaction act attractively.

\subsection{Flow equation of the continuous similarity transformation}

Below, we give all specific informations concerning the CST performed in the main body of the manuscript: the initial Hamiltonian $\mathcal{H}$ and the initial generator $\eta$, the flowing Hamiltonian $\mathcal{H}(\ell)$ and the flowing generator $\eta (\ell)$ as well as the corresponding flow equations.

\subsection{Initial Hamiltonian and initial generator}  
The initial Hamiltonian before the CST is given by 
 \begin{equation}
   \mathcal{H} = E_0 + \sum_{\k}\omega_{\k} 
	( \alpha^\dagger_{\k}\alpha^{\phantom{\dagger}}_{\k}+ 
	\beta^\dagger_{\k}\beta^{\phantom{\dagger}}_{\k})+\mathcal{V}\,.
\end{equation}
The coefficients of the quadratic part are \mbox{$E_0/N=-2J (S^2+AS +A^2/4)$}, \mbox{$\omega_{\k}=2J(2S+A)\sqrt{1-\gamma_{\k}^2}$} with \mbox{$\gamma_{\k}=(\cos k_x+\cos k_y)/2$}, \mbox{$A:=\frac{2}{N}\sum_\k(1-\sqrt{1-\gamma_{\k}^2})$}. The two-magnon interaction $\mathcal{V}$ is given explicitly by 
\bea
\mathcal{V} &=&  -  \frac {J}{N}\sum_{1234}
\dG{1+2-3-4}\; l_1l_2l_3l_4\nonumber\\
&&
\Big[V_{1234}^{(1)} \alpha_1^\dag \alpha_2^\dag \alpha_3^{\phantom{\dagger}} \alpha_4^{\phantom{\dagger}} 
+ 2V_{1234}^{(2)}\alpha_1^\dag \beta_{-2}^{\phantom{\dagger}}\alpha_3^{\phantom{\dagger}} \alpha_4^{\phantom{\dagger}} 
+ 2V_{1234}^{(3)}\alpha_1^\dag \alpha_2^\dag \beta_{-3}^\dag \alpha_{4}
+ 4V_{1234}^{(4)}\alpha_1^\dag \alpha_3^{\phantom{\dagger}} \beta_{-4}^\dag \beta_{-2}^{\phantom{\dagger}}
\non \\
&+& 2V_{1234}^{(5)}\beta_{-4}^\dag \alpha_3 \beta_{-2} \beta_{-1}^{\phantom{\dagger}}
+ 2V_{1234}^{(6)}\beta_{-4}^\dag \beta_{-3}^\dag \alpha_{2}^\dag \beta_{-1}^{\phantom{\dagger}} 
+ V_{1234}^{(7)}\alpha_1^\dag \alpha_2^\dag \beta_{-3}^\dag \beta_{-4}^\dag
+ V_{1234}^{(8)}\beta_{-1}^{\phantom{\dagger}} \beta_{-2}^{\phantom{\dagger}}\alpha_3^{\phantom{\dagger}} \alpha_4^{\phantom{\dagger}}
+V_{1234}^{(9)}\beta_{-4}^\dag \beta_{-3}^\dag \beta_{-2}^{\phantom{\dagger}} \beta_{-1}^{\phantom{\dagger}}\Big]\,.\quad 
\eea
The subscripts $i=1,2,3,4$ stand for the momenta ${\bf k}_i$ and  $-i$ stands for $-{\bf k}_i$.
The conservation of momentum in the lattice is ensured by the Kronecker symbol $\dG{k}$ which implies ${\bf k}=0$ modulo reciprocal lattice vectors from the reciprocal lattice $\Gamma_A^*$ of the $A$-sites, i.e., ${\bf g}\in\Gamma_A^*$ means
${\bf g}=(n\pi,m\pi)$ with the integers $n, m$ if the lattice constant of the original square lattice is set to unity. The vertex functions  $V_{1234}^{(i)}$ are given explicitly in Ref.~\onlinecite{uhrig13}.

\subsection{Flowing Hamiltonian and flowing generator}  

We include all terms in the flowing Hamiltonian $\mathcal{H}(\ell)$ with a scaling dimension $d\leq 2$. Explicitly, one finds
\begin{eqnarray} 
\mathcal{H}(\ell) = E_0 \ + 
&\sum_{\bf1}& \w{1} \left(\ad{1}\aoo{1}+\bd{1}\bo{1}\right)+\coeffgamma{1}\left(\ad{1}\bd{\uminus1}+\aoo{1}\bo{\uminus1}\right)
\non\\
      + &\sum_{\bf 1,2,3,4}& \Bigl\{\coeff{1}{1 2 3 4 } \ad{1}\ad{2}\aoo{3}\aoo{4}+
			\coeff{2}{1 2 3 4 } \ad{1}\aoo{2}\bd{3}\bo{4} + \coeff{3}{1 2 3 4 } 
			\bd{1}\bd{2}\bo{3}\bo{4} \Bigr.
\non\\
&+& \coeff{4}{1 2 3 4 } \ad{1}\ad{2}\aoo{3}\bd{4}  + 
			\coeff{5}{1 2 3 4 } \ad{1}\bd{2}\bd{3}\bo{4} + \coeff{6}{1 2 3 4 } 
			\ad{1}\aoo{2}\aoo{3}\bo{4}
\non\\
\Bigl. &+&\coeff{7}{1 2 3 4 } \aoo{1}\bd{2}\bo{3}\bo{4} +
					\coeff{8}{1 2 3 4 } \ad{1}\ad{2}\bd{3}\bd{4} + \coeff{9}{1 2 3 4 } 
					\aoo{1}\aoo{2}\bo{3}\bo{4} \Bigr\}
\quad.
\end{eqnarray}
The corresponding flowing generator is then given by 
\begin{eqnarray} 
\eta(\ell) =  
&\sum_{\bf1}&\coeffgamma{1}\left(\ad{1}\bd{\uminus1}-\aoo{1}\bo{\uminus1}\right)
\non\\
      +&\sum_{\bf 1,2,3,4}& \Bigl\{\coeff{4}{1 2 3 4 } \ad{1}\ad{2}\aoo{3}\bd{4}  + 
			\coeff{5}{1 2 3 4 } \ad{1}\bd{2}\bd{3}\bo{4} - \coeff{6}{1 2 3 4 } 
			\ad{1}\aoo{2}\aoo{3}\bo{4}
\non\\
\Bigl. 
&-&\coeff{7}{1 2 3 4 } \aoo{1}\bd{2}\bo{3}\bo{4} +
					\coeff{8}{1 2 3 4 } \ad{1}\ad{2}\bd{3}\bd{4} - \coeff{9}{1 2 3 4 } 
					\aoo{1}\aoo{2}\bo{3}\bo{4} \Bigr\}
\quad.
\end{eqnarray}

The coefficients $E_0$, $\w{1}$, $\coeffgamma{1}$, and $\coeff{i}{1 2 3 4 }$ depend on the flow parameter $\ell$ and satisfy the initial conditions
\begin{subequations}
\begin{eqnarray} 
  E_0\bigl|_{\ell=0}&=&-2J (S^2+AS +A^2/4)
\\
 \w{1}\bigl|_{\ell=0}&=& 2J(2S+A)\sqrt{1-\gamma_{\bf 1}^2}
\\
   \coeffgamma{1}\bigl|_{\ell=0}&=&0
\\
   \coeff{1}{1234}\bigl|_{\ell=0}&=& -  {l_1l_2l_3l_4}
	\frac {J}{N} V_{1234}^{(1)} \dG{1+2-3-4} 
\\
   \coeff{2}{1234}\bigl|_{\ell=0}&=& -  4{l_1l_2l_3l_4}
	\frac {J}{N}V_{1\uminus42\uminus3}^{(4)} \dG{1-2+3-4}
\\
   \coeff{3}{1234}\bigl|_{\ell=0}&=& -  {l_1l_2l_3l_4}
	\frac {J}{N}V_{\uminus4\uminus3\uminus2\uminus1}^{(9)} \dG{1+2-3-4}      
\\
   \coeff{4}{1234}\bigl|_{\ell=0}&=& -  2{l_1l_2l_3l_4}
	\frac {J}{N}V_{1 2\uminus4 3}^{(3)} \dG{1+2-3+4}
\\
   \coeff{5}{1234}\bigl|_{\ell=0}&=& -  2{l_1l_2l_3l_4}
	\frac {J}{N}V_{\uminus4 \uminus1\uminus2\uminus3}^{(6)} \dG{1+2+3-4}
\\
   \coeff{6}{1234}\bigl|_{\ell=0}&=& -  2{l_1l_2l_3l_4}
	\frac {J}{N}V_{1\uminus4 2 3}^{(2)} \dG{1-2-3-4}      
\\
   \coeff{7}{1234}\bigl|_{\ell=0}&=& -  2{l_1l_2l_3l_4}
	\frac {J}{N}V_{\uminus4\uminus3 1\uminus2}^{(5)} \dG{-1+2-3-4}
\\
   \coeff{8}{1234}\bigl|_{\ell=0}&=& -  {l_1l_2l_3l_4}
	\frac {J}{N}V_{ 1 2\uminus3\uminus4}^{(7)} \dG{1+2+3+4}
\\
   \coeff{9}{1234}\bigl|_{\ell=0}&=& -  {l_1l_2l_3l_4}
	\frac {J}{N}V_{\uminus3\uminus4 1 2}^{(8)} \dG{-1-2-3-4}\quad .
\end{eqnarray}
\label{eq:initialcondition}
\end{subequations}

Note that the hexatic three-magnon part of the Hamiltonian consisting of six magnon annihilation or creation operators has a scaling dimension $d=4$ and are therefore not relevant at studied level of truncation $d=2$.

\subsection{Flow equations}  
Inserting $\mathcal{H}(\ell)$ and $\eta (\ell)$ into the flow equation \mbox{$\partial_{\ell} \mathcal{H}(\ell) = \left[\eta(\ell),\mathcal{H}(\ell) \right]$} and keeping self-similarly all operators already present in $\mathcal{H}(\ell)$, one obtains the following flow equations
\begin{subequations}
\begin{eqnarray}                   
\partial_l E_0 =  &\sum_{\bf1,2 3,4}&(-8)\coeff{8}{1 2 3 4 }\coeff{9}{1 2 3 4 }   + \sum_{\bf1}(-2)\coeffgamma{1}\coeffgamma{1}  
\\   
\partial_l\w{1}     =  (-2)\coeffgamma{1}\coeffgamma{1} + &\sum_{\bf 3,4}& \Bigl\{  -4\coeff{4}{3 4 1 5}  \coeff{6}{1 3 4 5}\dG{3+4-1+5}\Bigr.  
\non \\
&\phantom{\sum_{\bf 3,4}}&        -16 \coeff{8}{1 3 4 5 }\coeff{9}{1 3 4 5 }\dG{3+4+1+5}
\non \\
\Bigl.  &\phantom{\sum_{\bf 3,4}}&      -4\coeff{8}{1 3 1 {\uminus3} }\coeffgamma{3}   -4\coeff{6}{1 1 3 {\uminus3} }\coeffgamma{3} \Bigr\} 
\\ 
\partial_l \coeffgamma{1}= -2\coeffgamma{1}\w{1} -8 
&\sum_{3,4,5}& \Bigl\{ \coeff{4}{3415}\coeff{9}{34\uminus15}\dG{-1+3+4+5}\Bigr.
\non \\
&\phantom{=}& \Bigl.-8\coeff{5}{345\uminus1}\coeff{9}{1345} \dG{1+3+4+5}\Bigr. 
\non \\
&\sum_{3}& \Bigl\{ \coeff{2}{13\uminus1\uminus3}\coeffgamma{3}\Bigr.
\non \\
 &\phantom{=}& \Bigl.\coeff{8}{13\uminus1\uminus3}\coeffgamma{3}\Bigr\}
\end{eqnarray}
\end{subequations}

\begin{subequations}
\begin{eqnarray}  
 \partial_l  \coeff{1}{1 2 3 4 }  &=&  \dG{1+2-3-4}\Bigl\{(-1) \coeff{4}{1 2 4 3 }\coeffgamma{3} \Bigr.  \nonumber\\ 
					&\phantom{=}& 		(-1) \coeff{4}{1 2 3 4 }\coeffgamma{4}\nonumber\\ 	
					&\phantom{=}& 		(-1) \coeff{6}{2 3 4 5 }\coeffgamma{1}\nonumber\\ 
					&\phantom{=}& 		(-1) \coeff{6}{1 3 4 5 }\coeffgamma{2}\nonumber\\ 
					 &\sum_{\bf 5,6}& (-4) \coeff{4}{2 5 4 6 }\coeff{6}{1 3 5 6 }\dG{2+5-4+6} \nonumber\\ 
					&\phantom{=}& 		(-4) \coeff{4}{1 5 3 6 }\coeff{6}{2 4 5 6 }\dG{1+5-3+6} \nonumber\\ 
					&\phantom{=}&  \Bigl. 	(-4) \coeff{8}{1 2 5 6 }\coeff{9}{3 4 5 6 }\dG{1+2+5+6}\Bigr\} \nonumber\\ 
\end{eqnarray}

\begin{eqnarray}  
 \partial_l  \coeff{2}{1 2 3 4 }  &=&  \dG{1-2+3-4}\Bigl\{(-1) \coeff{4}{1 \uminus4 2 3 }\coeffgamma{\uminus4} \Bigr. 
\non\\ 
&\phantom{=}& 	(-1) \coeff{5}{1 3 \uminus2 4 }\coeffgamma{\uminus2}
\non\\ 	
&\phantom{=}& 	(-1) \coeff{6}{1 2 \uminus3 4 }\coeffgamma{\uminus3}
\non\\ 
&\phantom{=}& 	(-1) \coeff{7}{2 3 4 \uminus1 }\coeffgamma{\uminus1}
\non\\ 
&\sum_{\bf 5,6}& (-4) \coeff{4}{5 6 2 3 }\coeff{6}{1 5 6 4 }\dG{5+6-2+3} 
\non\\ 
&\phantom{=}& 		(-4) \coeff{5}{1 5 6 4 }\coeff{7}{2 3 5 6 }\dG{1+5+6-4} 
\non\\ 
&\phantom{=}&            (-8) \coeff{4}{1 5 2 6 }\coeff{6}{5 3 4 6 }\dG{1+5-2+6} 
\non\\ 
&\phantom{=}& 		(-8) \coeff{5}{5 3 6 4 }\coeff{7}{1 2 5 6 }\dG{5+3+6-4} 
\non\\ 
&\phantom{=}&  \Bigl. 	(-32) \coeff{8}{1 5 3 6 }\coeff{9}{2 5 4 6 }\dG{1+5+3+6}\Bigr\} 
\end{eqnarray}

\begin{eqnarray}  
\partial_l \coeff{3}{1 2 3 4 }  &=&    \dG{1+2-3-4}   \Bigl\{   (-1) \coeff{5}{{\uminus3} 1 2 4 }\coeffgamma{\uminus3} \Bigr. 
\non\\   
&\phantom{=}&          (-1) \coeff{5}{\uminus4 1 2 3 }\coeffgamma{\uminus4 }	 
\non\\ 
&\phantom{=}&          (-1) \coeff{7}{\uminus2 1 3 4 }\coeffgamma{\uminus2 } 
\non\\ 
&\phantom{=}&          (-1) \coeff{7}{{\uminus1} 2 3 4 }\coeffgamma{\uminus1 }  
\non\\ 
&\sum_{\bf5}& (-4) \coeff{5}{5 2 6 4 }\coeff{7}{5 1 3 6 }\dG{5+1+6-4} 
\non\\ 
&\phantom{=}&            (-4) \coeff{5}{5 1 6 3 }\coeff{7}{5 2 4 6 }\dG{5+1+6-3}   
\non\\ 
&\phantom{=}&   \Bigl. (-4) \coeff{8}{5 6 1 2 }\coeff{9}{5 6 3 4 }\dG{5+6+1+2} \Bigr\}
\end{eqnarray}

\begin{eqnarray}    
\partial_l \coeff{4}{1 2 3 4 }  &=&   \dG{1+2-3+4} \Bigl\{ \left(\w{3} - \w{1} - \w{2} - \w{4}\right) \coeff{4}{1 2 3 4 } \Bigr.	
\non\\ 
&\phantom{=}&         (-2)\coeff{1}{123\uminus4}\coeffgamma{\uminus4} 
\non\\    
&\phantom{=}&            \left(-\frac{1}{2}\right)\left(\coeff{2}{234\uminus1}\coeffgamma{1} +\coeff{2}{134\uminus2}\coeffgamma{2}\right) 
\non\\      
&\phantom{=}&                (-4)\coeff{8}{124\uminus3}\coeffgamma{3}  
\non\\                   		    
&\sum_{\bf 5,6}& (-2) \coeff{1}{1 2 5 6 }\coeff{4}{5 6 3 4 }\dG{1+2-5-6}   
\non\\  
&\phantom{=}&                        (-1) \coeff{2}{2 5 4 6 }\coeff{4}{1 5 3 6 }\dG{2-5+4-6}  
\non\\ 
&\phantom{=}&                        (-1) \coeff{2}{1 5 4 6 }\coeff{4}{2 5 3 6 }\dG{1-5+4-6}    
\non\\ 
          &\phantom{=}&                        (-8) \coeff{6}{2 3 5 6 }\coeff{8}{1 5 4 6 }\dG{2-3-5-6}    
\non\\                                         
          &\phantom{=}&                        (-8) \coeff{6}{1 3 5 6 }\coeff{8}{2 5 4 6 }\dG{1-3-5-6}    
\non\\  
          &\phantom{=}&                 \Bigl. (-4) \coeff{7}{3 4 5 6 }\coeff{8}{1 2 5 6 }\dG{3-4+5+6} \Bigr\}  
\end{eqnarray}

\begin{eqnarray}  
    \partial_l \coeff{5}{1 2 3 4 }  &=& \dG{1+2+3-4} \Bigl\{  \left(-\w{1} - \w{2} - \w{3} + \w{4}\right) \coeff{5}{1 2 3 4 }	\Bigr. 
\non\\
&\phantom{=}&         (-2)\coeff{3}{234\uminus1}\coeffgamma{1} 
\non\\    
&\phantom{=}&            \left(-\frac{1}{2}\right)\left(\coeff{2}{1\uminus234}\coeffgamma{\uminus2} +\coeff{2}{1\uminus324}\coeffgamma{\uminus3}\right)
 \non\\      
&\phantom{=}&                (-4)\coeff{8}{1\uminus423}\coeffgamma{\uminus4}  
\non\\ 	     		    
&\sum_{\bf5,6}& (-2) \coeff{1}{5 6 2 3 }\coeff{6}{1 5 6 4 }\dG{5+6-2-3}     
\non\\
&\phantom{=}&                   (-1) \coeff{2}{1 5 3 6 }\coeff{5}{5 2 6 4 }\dG{1-5+3-6}     
\non\\
&\phantom{=}&                  (-1) \coeff{2}{1 5 2 6 }\coeff{5}{5 3 6 4 }\dG{1-5+2-6}     
\non\\
&\phantom{=}&                   (-8) \coeff{7}{5 3 4 6 }\coeff{8}{1 5 2 6 }\dG{5-3+4+6}     
\non\\                                       
&\phantom{=}&                    (-8) \coeff{7}{5 2 4 6 }\coeff{8}{1 5 3 6 }\dG{5-2+4+6}     
\non\\
&\phantom{=}&               \Bigl. (-4) \coeff{6}{1 5 6 4 }\coeff{8}{5 6 2 3 }\dG{1-5-6-4} \Bigr\}
\end{eqnarray}

\begin{eqnarray}  
\partial_l \coeff{6}{1 2 3 4 }  &=& \dG{1-2-3-4}\Bigl\{ \left(\w{1} - \w{2} - \w{3}- \w{4}\right) \coeff{6}{1 2 3 4 }	\Bigr.	  
\non\\ 
        &\phantom{=}&         (-2)\coeff{1}{1\uminus423}\coeffgamma{\uminus4} 
\non\\    
        &\phantom{=}&            \left(-\frac{1}{2}\right)\left(\coeff{2}{13\uminus24}\coeffgamma{2} +\coeff{2}{12\uminus24}\coeffgamma{3}\right) 
\non\\      
        &\phantom{=}&                (-4)\coeff{9}{234\uminus1}\coeffgamma{1}  
\non\\ 	     			        		    
                                &\sum_{\bf5,6}& (-2) \coeff{1}{5 6 2 3 }\coeff{6}{1 5 6 4 }\dG{5+6-2-3}    
\non\\ 
                                     &\phantom{=}&        (-1) \coeff{2}{5 3 6 4 }\coeff{6}{1 2 5 6 }\dG{5-3+6-4} 
\non\\
                                     &\phantom{=}&        (-1) \coeff{2}{5 2 6 4 }\coeff{6}{1 3 5 6 }\dG{5-2+6-4} 
\non\\
                                     &\phantom{=}&        (-8) \coeff{4}{1 5 3 6 }\coeff{9}{2 5 4 6 }\dG{1+5-3+6} 
 \non\\                                          
                                     &\phantom{=}&        (-8) \coeff{4}{1 5 2 6 }\coeff{9}{3 5 4 6 }\dG{1+5-2+6}   
\non\\  
                                      &\phantom{=}&      \Bigl. (-4) \coeff{5}{1 5 6 4 }\coeff{9}{2 3 5 6 }\dG{1+5+6-4}  \Bigr\}
\end{eqnarray}

\begin{eqnarray}  
\partial_l \coeff{7}{1 2 3 4 }  &=& \dG{1-2+3+4} \Bigl\{   \left(-\w{1} + \w{2} - \w{3} - \w{4}\right) \coeff{7}{1 2 3 4 }\Bigr. 
\non\\	
&\phantom{=}&         (-2)\coeff{3}{2\uminus134}\coeffgamma{1} 
\non\\    
&\phantom{=}&            \left(-\frac{1}{2}\right)\left(\coeff{2}{\uminus3124}\coeffgamma{\uminus3} +\coeff{2}{\uminus4123}\coeffgamma{\uminus4}\right) 
\non\\      
&\phantom{=}&                (-4)\coeff{9}{1\uminus234}\coeffgamma{\uminus2}  
\non\\ 	     		     		    
&\sum_{\bf5,6}& (-2) \coeff{3}{5 6 3 4 }\coeff{7}{1 2 5 6 }\dG{5+6-3-4} +  
\non\\  
&\phantom{=}&             (-1) \coeff{2}{5 1 6 4 }\coeff{7}{5 2 3 6 }\dG{5-1+6-4} 
\non\\
&\phantom{=}&             (-1) \coeff{2}{5 1 6 3 }\coeff{7}{5 2 4 6 }\dG{5-1+6-3} 
\non\\
&\phantom{=}&             (-8) \coeff{5}{5 2 6 4 }\coeff{9}{1 5 3 6 }\dG{5+2+6-4}    
\non\\                                        
&\phantom{=}&             (-8) \coeff{5}{5 2 6 3 }\coeff{9}{1 5 4 6 }\dG{5+2+6-3}   
\non\\  
&\phantom{=}&           \Bigl.(-4) \coeff{4}{5 6 1 2 }\coeff{9}{5 6 3 4 }\dG{5+6-1+2}  \Bigr\}
\end{eqnarray}

\begin{eqnarray}     
   \partial_l \coeff{8}{1 2 3 4 } &=& \dG{1+2+3+4}\Bigl\{  \left(-\w{1} - \w{2} - \w{3} - \w{4}\right) \coeff{8}{1 2 3 4 }\Bigr. \non\\			     		    
& \sum_{\bf5,6}& (-2) \coeff{1}{1 2 5 6 }\coeff{8}{5 6 3 4 }\dG{1+2-5-6} 
\non\\
&\phantom{=}&             (-1) \coeff{2}{2 5 4 6 }\coeff{8}{1 5 3 6 }\dG{2-5+4-6} 
\non\\
&\phantom{=}&            (-1) \coeff{2}{1 5 4 6 }\coeff{8}{2 5 3 6 }\dG{1-5+4-6} 
\non\\
&\phantom{=}&            (-1) \coeff{2}{2 5 3 6 }\coeff{8}{1 5 4 6 }\dG{2-5+3-6}        
\non\\                                    
&\phantom{=}&            (-1) \coeff{2}{1 5 3 6 }\coeff{8}{2 5 4 6 }\dG{1-5+3-6}     
\non\\ 
&\phantom{=}&        \Bigl. (-2) \coeff{3}{3 4 5 6 }\coeff{8}{1 2 5 6 }\dG{3+4-5-6}  \Bigr\}
\end{eqnarray}

\begin{eqnarray}  
\partial_l \coeff{9}{1 2 3 4 }  &=& \dG{1+2+3+4} \Bigl\{\left(-\w{1} - \w{2} - \w{3} - \w{4}\right) \coeff{9}{1 2 3 4 }\Bigr.
\non\\		     		    
&\sum_{\bf 5,6}& (-2) (-2) \coeff{1}{5 6 1 2 }\coeff{9}{5 6 3 4 }\dG{5+6-1-2}    
\non\\ 
&\phantom{=}&            (-1) \coeff{2}{5 2 6 4 }\coeff{9}{1 5 3 6 }\dG{5-2+6-4} 
\non\\
&\phantom{=}&            (-1) \coeff{2}{5 1 6 4 }\coeff{9}{2 5 3 6 }\dG{5-1+6-4} 
\non\\
&\phantom{=}&    (-1) \coeff{2}{5 2 6 3 }\coeff{9}{1 5 4 6 }\dG{5-2+6-3}     
\non\\                                       
&\phantom{=}&   (-1) \coeff{2}{5 1 6 3 }\coeff{9}{2 5 4 6 }\dG{5-1+6-3}    
\non\\ 
&\phantom{=}&   \Bigl.   (-2) \coeff{3}{5 6 3 4 }\coeff{9}{1 2 5 6 }\dG{3+4-5-6}  \Bigr\}
\end{eqnarray}

\end{subequations}
Note that the Kronecker symbols $\dG{k}$ in the sums are redundant since the coefficients intrinsically conserve the total momentum by definition (see Eq.~\eqref{eq:initialcondition}). They are included to underline momentum conservation.

\end{document}